\documentclass[12pt,a4paper]{article}
\usepackage{graphicx}
\begin{document}
\textwidth=135mm
 \textheight=200mm
\begin{center}
{\bfseries \Large High-Energy Neutrino Astronomy:} 
\vskip 2mm
{\bfseries \Large Where do we stand, where do we go?}
\vskip 5mm
Christian Spiering$^{\dag}$ 
\vskip 4mm
{\small {\it $^\dag$ DESY, Platanenallee 6 D\,15738 Zeuthen, Germany}}
\\
{email:\,\texttt{christian.spiering@desy.de}
}
\end{center}
\vskip 5mm
\centerline{\bf Abstract}
{\small With the identification of a diffuse flux of 
astrophysical (``cosmic'') neutrinos in the TeV-PeV energy range, 
IceCube has opened a new window to the Universe. However, the corresponding cosmic landscape is still uncharted: so far, the observed flux does not show any clear association with known source classes. In the present talk, I sketch the way from Baikal-NT200 to IceCube and summarize IceCube's recent astrophysics results. Finally, I describe the present projects to build even larger detectors:
GVD in Lake Baikal, KM3NeT in the Mediterranean Sea and IceCube-Gen2 at the South Pole. These detectors will allow studying the high-energy neutrino sky in much more detail than the present arrays permit.}
\vskip 8mm

\section{Introduction}

The first conceptual ideas how to detect high energy neutrinos date back to the late fifties. 
The long evolution towards detectors with a realistic discovery potential
started in the seventies, by the pioneering works 
in the Pacific Ocean close to Hawaii (DUMAND).
The DUMAND 1978 design envisaged an array of about 20\,000 photomultipliers
spread over a 1.26 cubic kilometer volume of water.  
This project was terminated in 1995, but the baton was
taken by the projects NT200 in Lake Baikal, AMANDA at the South Pole,
ANTARES in the Mediterranean Sea and, again at the South Pole, IceCube (see for detailed
information on the history and on corresponding references \cite{Spiering-History}).
But only now, half a century after the first concepts, a cubic kilometer detector 
is in operation: IceCube at the South Pole.
With the discovery of a flux of high-energy neutrinos of astrophysical origin (``cosmic neutrinos'')
in 2013 \cite{Science-2013}, the IceCube Neutrino Observatory has opened a new window to the Universe of non-thermal cosmic processes. 
A next generation of arrays is under construction or planned: KM3NeT in the
Mediterranean Sea \cite{KM3NeT}, the Gigaton Volume Detector GVD in Lake Baikal
\cite{Baikal}, 
and IceCube-Gen2 \cite{Gen2}.

The primary goal of these detectors is identifying the sources of high-energy cosmic rays. 
In contrast to charged particles, neutrinos are not deflected in cosmic magnetic fields and keep their direction; in contrast to gamma rays they provide a direct, water-tight prove for the acceleration of hadrons in the emitting sources. This makes them unique tracers of sources of cosmic rays.
On the other hand, due to their small interaction cross section they are difficult to detect: The         ``neutrino effective area'' of the 1 km$^3$ IceCube detector (essentially the geometrical area multiplied with the interaction probability, the trigger efficiency and the transparency to neutrinos of the Earth)
is less than 1\,m$^2$ at 1\,TeV and of the order of 100\,m$^2$ at 100\,TeV \cite{point-sources}. It is therefore no surprise that it took several decades to detect cosmic neutrinos. 

Neutrino telescopes are multi-purpose detectors. Apart from investigating cosmic neutrinos, they exploit atmospheric neutrinos to study neutrino oscillation, to search for sterile neutrinos or to test fundamental laws of physics. They are used to search for neutrinos from Dark Matter annihilations in the Sun or the Galactic halo, to search for exotic particles like magnetic monopoles, or to study muons from cosmic-ray induced air showers. 

This paper focuses to the search for neutrinos from cosmic acceleration processes. I will further focus to the developments at Lake Baikal and at the South Pole where I have been involved myself over a long period.

\section{From Baikal NT200 to IceCube}

\subsection{NT200 and AMANDA}

\begin{figure}[ht]
\includegraphics[width=0.80\linewidth]{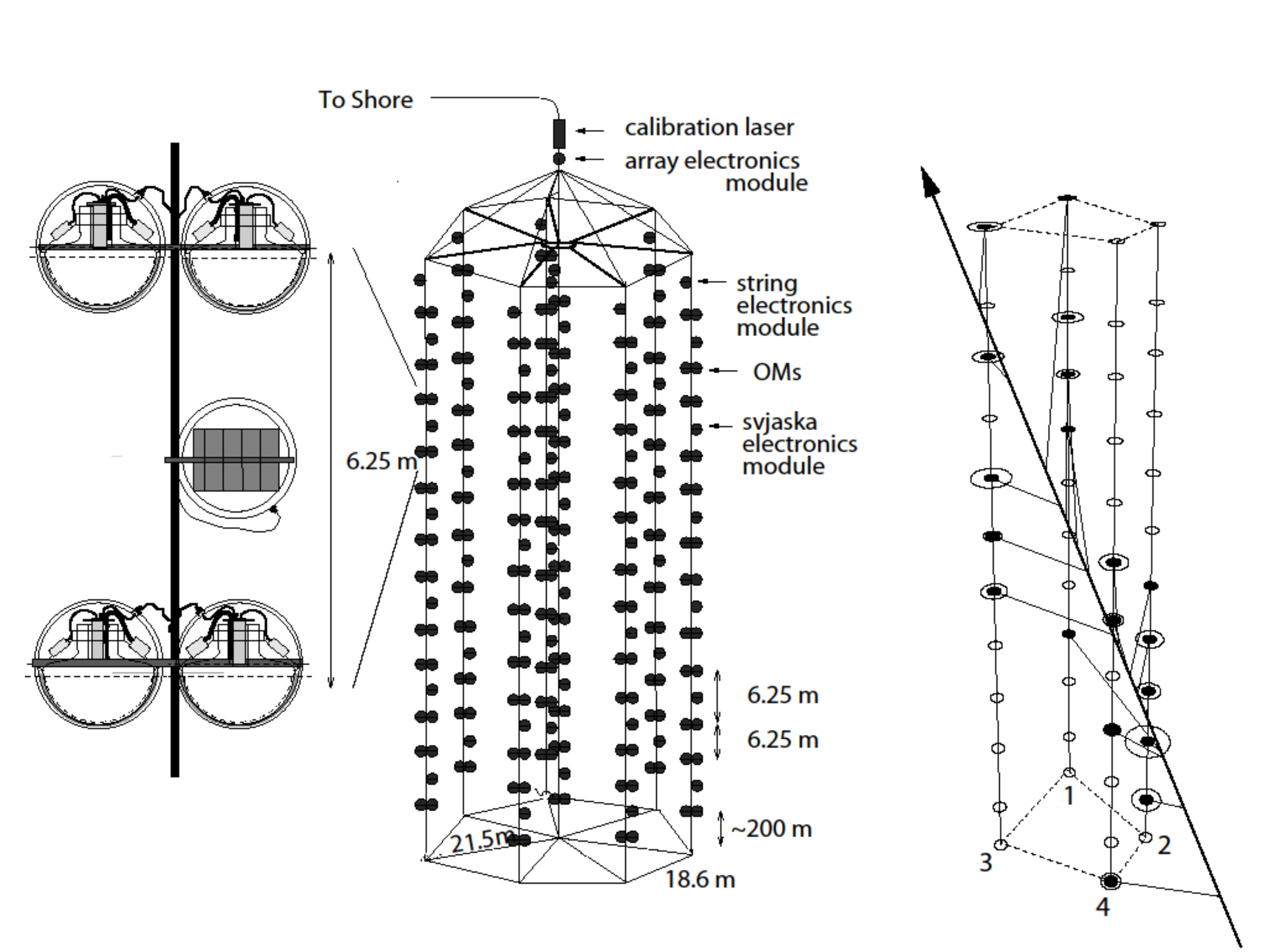}
\caption{\small {\it Left}: The Baikal Neutrino Telescope NT200. 
{\it Right}: One of the first upward moving
muons from a neutrino interaction recorded with the 4-string stage of the
detector in 1996 \cite{Baikal-atm-Balkanov-1999}. The Cherenkov light from the
muon is recorded by 19 channels.}
\label{nt200}
\end{figure}

First test deployments in Lake Baikal started in 1981.
The construction of the NT200 detector was started in 1993,
about 30 km South-West from the outflow of Lake Baikal into
the Angara river, at a distance of
3.6\,km to shore and at a depth of about 1.1\,km, and was completed in 1998.
NT200 was an array of 192 optical modules 
(glass spheres containing large photomultipliers) 
at eight strings, 72\,m in height and 43\,m in diameter 
(see Fig.\,\ref{nt200}). Actually,
this is not much more than twice the size of Super-Kamiokande. First
upward going muons (i.e.\,neutrino events) were found already with the
3-string version from 1993, and then with the 4-string version from 1996: 
the first ``underwater neutrinos'' ever! This was the first proof of principle
to detect neutrinos in open media and a breakthrough for the field.

\begin{figure}[ht]
\centering
\includegraphics[width=0.65\linewidth]{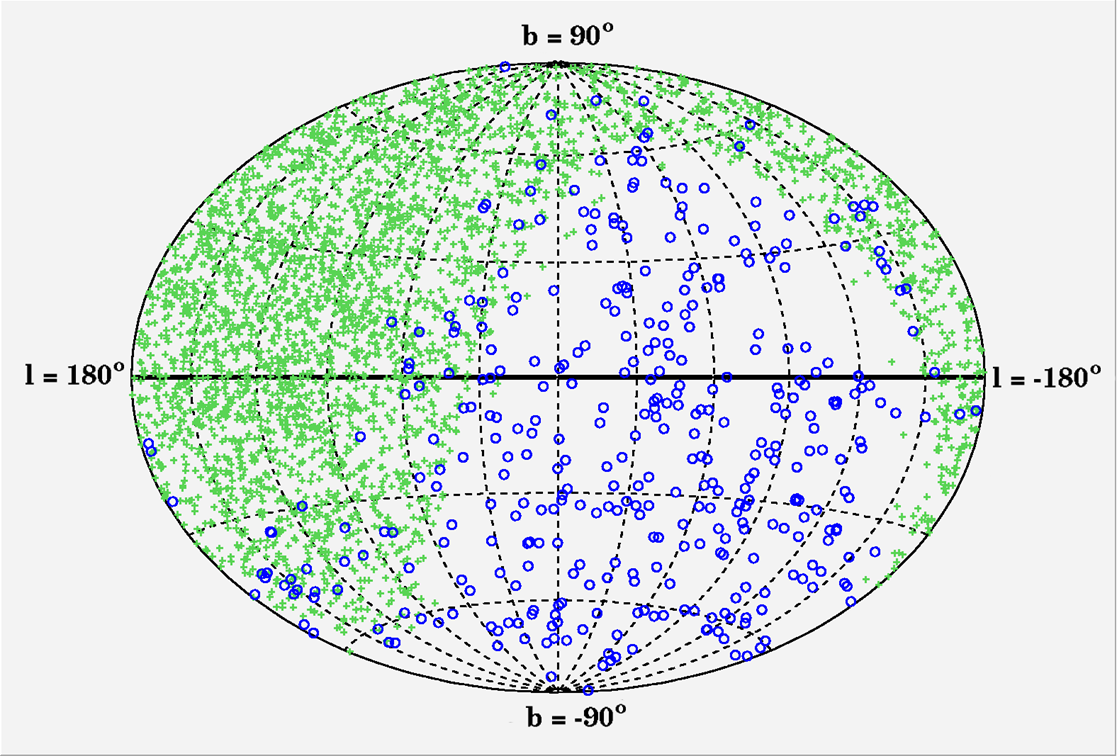}
\caption{\small A skyplot in Galactic coordinates, combining neutrinos detected by NT200 (circles) and AMANDA (crosses).
Compiled in 2005 \cite{Igor-Christian-skyplot}.
}
\label{A-B-skyplot}
\end{figure}

In 1988, a new, spectacular idea appeared on stage: to use Antarctic ice instead
of water as target and as detector medium.
The project was named AMANDA (Antarctic Muon And Neutrino Detection Array).
AMANDA was deployed some hundred meters from the Amundsen-Scott station, first
at a too shallow depth (where bubbles disturb light propagation), and then
from 1996 -- 2000 at depths between 1500 and 2000\,m. It consisted of
677 optical modules at 19 strings.
AMANDA was switched off in April 2009, after more than 9 years of data
taking in its full configuration and with 6959 neutrino events collected. Naturally
this sample was dominated by atmospheric neutrinos.  
No indication of point sources was found, and no excess of high-energy
events which might have pointed to an admixture of cosmic neutrinos.
Figure \ref{A-B-skyplot} 
shows -- as a kind of sentimental reminiscence -- a combination of NT200 and AMANDA
data compiled in 2005 (including all NT200 data and 2 years AMANDA data).

AMANDA provided record limits on fluxes for cosmic neutrinos, 
be it for diffuse fluxes (where the much smaller NT200 could compete for some years)
and for point sources -- steady as well as transient. 
AMANDA also extended the measured spectrum of atmospheric neutrinos by nearly
two orders of magnitude, from a few TeV to 200\,TeV.

\subsection{Mediterranean Projects}

First site studies in the Mediterranean Sea were performed in 1989,
leading to the NESTOR project, with the goal to install towers
of hexagonal floors close to Pylos in Greece. The first floor was deployed only 15 years
later, in 2004.
But the project was further and further delayed and the community split into
a Greek project (NESTOR), an Italy-based project (NEMO) and a
project in France (ANTARES). Only the French site made it to
a working -- and actually excellently working! -- detector.

The construction of ANTARES started in 2002 with the deployment of a shore cable.
The detector in its final 12-string configuration was
installed in 2006--2008 and has been operational since then.
The strings have lateral distances of 60--70\,m, and each of them carries
25\,triplets of optical modules at depths of 2.1--2.4\,km.
 
ANTARES has demonstrated that a stable operation of a deep-sea detector is possible.
Similar in size to AMANDA,  it has collected more than 8000 upward-going
muon tracks over eight years of operation.
With its excellent view of the Galactic plane and
good angular resolution, the telescope could constrain the Galactic origin of the 
cosmic neutrino flux reported by IceCube. ANTARES has explored he Southern sky and in particular
central regions of our Galaxy searching for point-like objects,
for extended regions of emission, and for signals from transient objects selected
through multi-messenger observations
\cite{Antares-results}.

\subsection{IceCube}

Like AMANDA, the IceCube Observatory \cite{IceCube} is located at the geographical South Pole. It consists of the main IceCube array with its subarray DeepCore and the surface array IceTop. IceCube comprises 5160 digital optical modules (DOMs) installed on 86 strings at ice
depths of 1450 to 2450\,m and covers 1 km$^3$ of ice. 
A string carries 60~DOMs. DeepCore, a high-density sub-array of eight strings 
at the center of IceCube, has smaller spacing and DOMs with more sensitive photomultipliers than
IceCube and sits in the midst of the clearest ice layers. This results in a
threshold of about 10\,GeV and opens a new venue for oscillation physics.
The threshold of the full IceCube detector is about 100\,GeV. In its final configuration,
IceCube takes data since spring 2011, with a duty cycle of more than 99\%. It collects about $10^5$ clean neutrino events per year, with nearly
99.9\% of them being of atmospheric origin.

\section{Where do we stand?}

\subsection{Diffuse Fluxes}

It has been predicted since long that the first evidence for extragalactic cosmic neutrinos would be provided by a diffuse flux rather than by single-source signals \cite{Lipari}. 
The first tantalizing hint to cosmic neutrinos in IceCube came from two shower-like events with energies 
$\approx 1$ PeV, discovered in 2012 and dubbed ``Ernie'' and ``Bert'' \cite{Ernie}. 
A follow-up search of the same data (May 2010 to April 2012) with a lowered threshold (30 TeV) 
provided 25 additional events. This analysis used only events starting in a fiducial volume of 
about 0.4 km$^3$ (High Energy Starting Events, or ``HESE''), using the other 60\% of IceCube as veto against all sorts of background. Energy spectrum and zenith angle distribution
of the 27 events excluded an only-atmospheric origin with $4.1\sigma$ and suggested that 
about 60\% were of cosmic origin, 
at energies above 100 TeV even about 80\% \cite{Science-2013}. A four-year data set with
54 neutrinos provided another shower-like PeV event (deposited energy $\approx$ 2 PeV) and
confirmed a dominant cosmic contribution with nearly $6.5\sigma$. Very recently, the results from 
a six-year sample have been presented \cite{ICRC-Kopper}, with 82 events above 30 TeV.
Figure 3
shows the  energies deposited by these events inside IceCube. 

\begin{figure}[h]
\vspace{-0.1cm}
\centering
\includegraphics[width=0.5\textwidth]{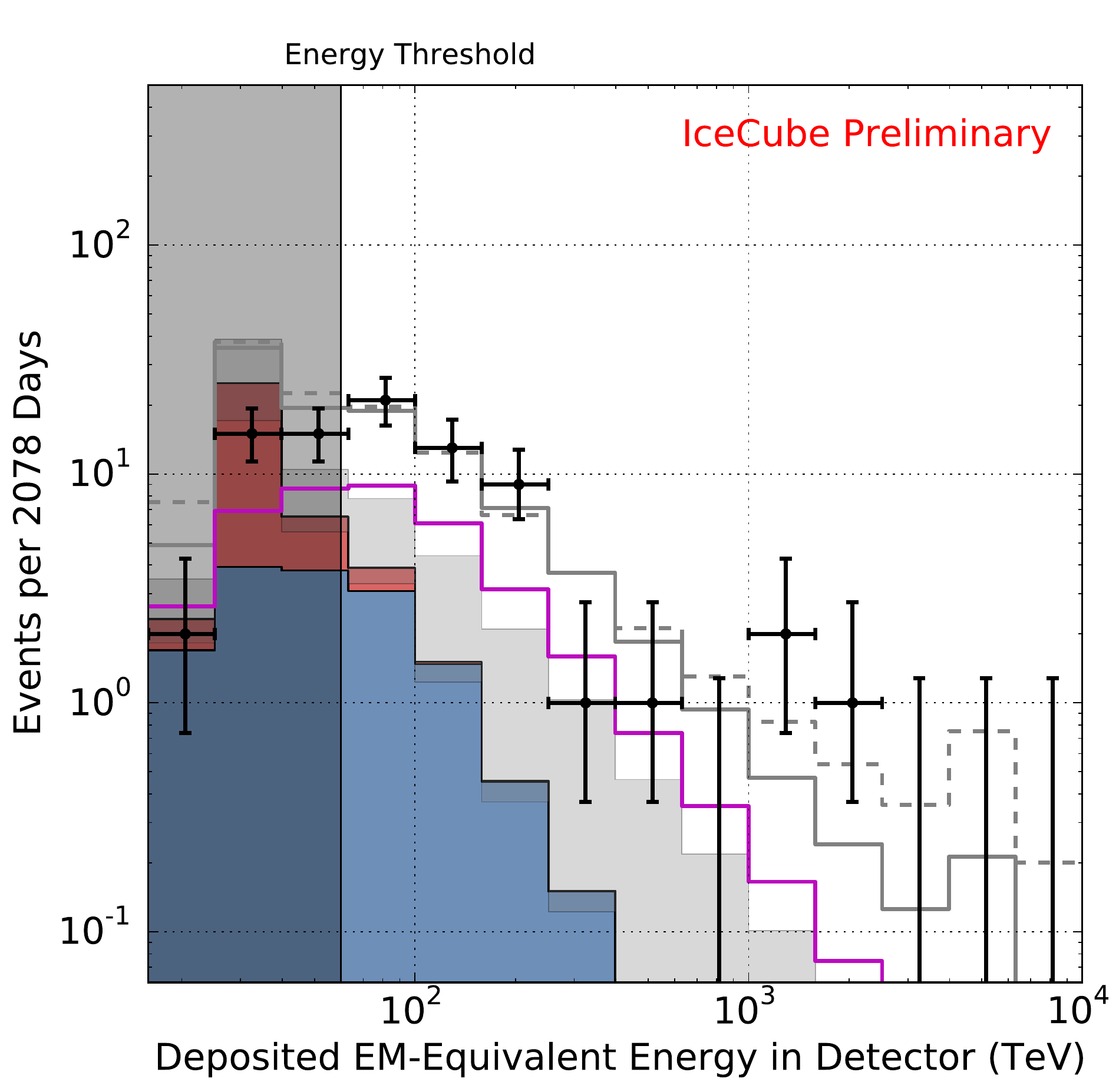}
  \caption{\small
	Distribution of the energy deposited by 82 events from the six-year HESE analysis. 
	Backgrounds of atmospheric origin 
come from punch-through down-going muons and from atmospheric neutrinos. 
While the flux of neutrinos from $\pi$ and K decays is well known (blue region), the neutrino flux 
from charm decays in the atmosphere is uncertain and dominates
the uncertainty of all background sources (gray region with 1$\sigma$ uncertainties). The best-fit astrophysical spectra are shown as gray lines, for a single power-law spectrum as solid line, for a two power-law model as dashed line. See \cite{ICRC-Kopper} for details.
}
\label{l-HESE-spectrum}
\end{figure}

A $5.6\sigma$ excess of high-energy cosmic neutrinos is also seen in the spectrum of secondary muons generated by neutrinos that have traversed the Earth, with a zenith angle less than 5 degrees above the horizon
(``upward through-going muons''\cite{muon-7years}).
Figure 4
shows the median neutrino energy. It is calculated for each energy deposited by the muon in the detector, assuming the best-fit spectrum. The highest energy muon has deposited $2.6\pm 0.3$ PeV inside the instrumented volume, which corresponds to a most probable neutrino energy of about 9 PeV.

\begin{figure}[ht]
\vspace{-0.1cm}
\centering
\includegraphics[width=0.65\textwidth]{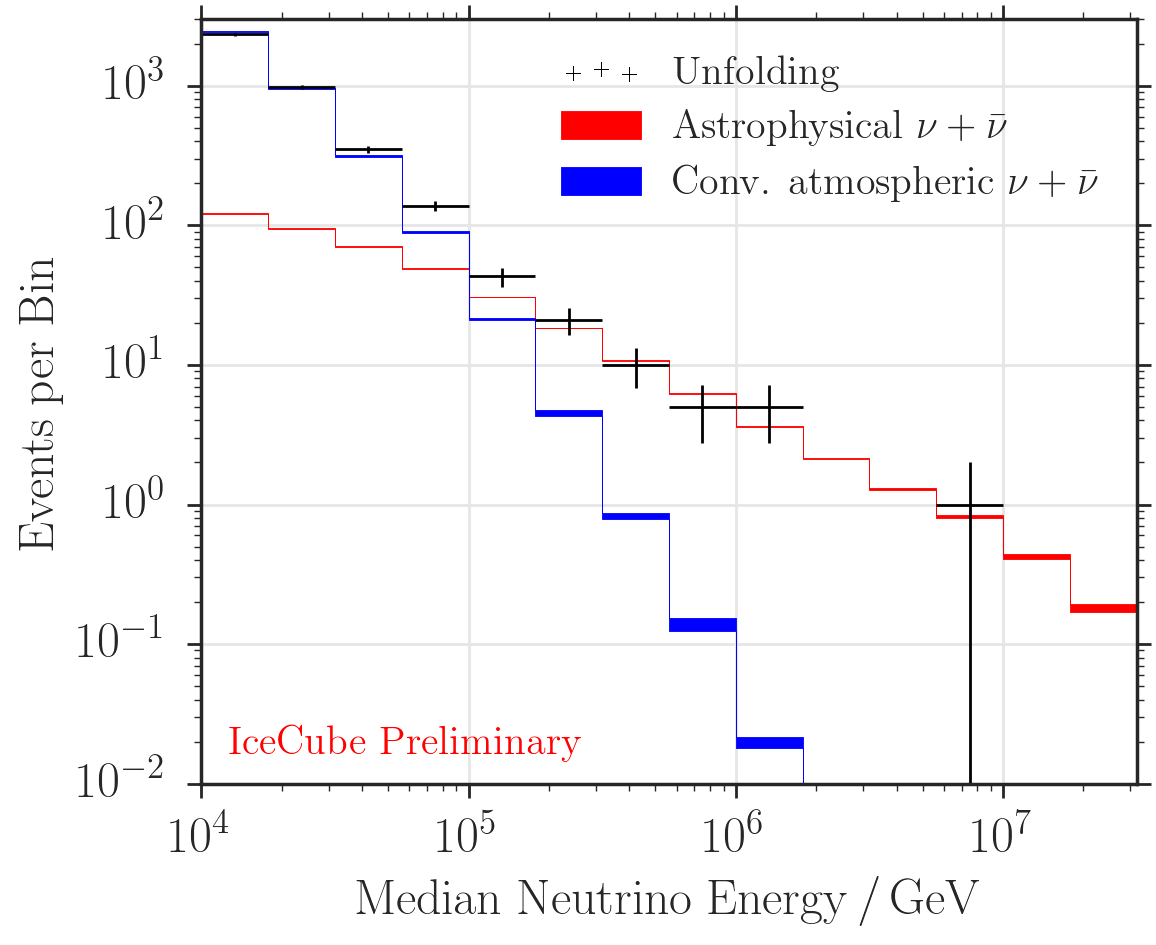}
  \caption{\small 
Spectrum of the median neutrino energy derived from the energy deposit of through-going muons with zenith angles less than 5 degrees above horizon (8 years sample \cite{muon-8years}).
}
\label{l-muon-energy-spectrum}
\vspace{-0.1cm}
\end{figure}

\begin{figure} 
\centering
\includegraphics[width=0.7\textwidth]{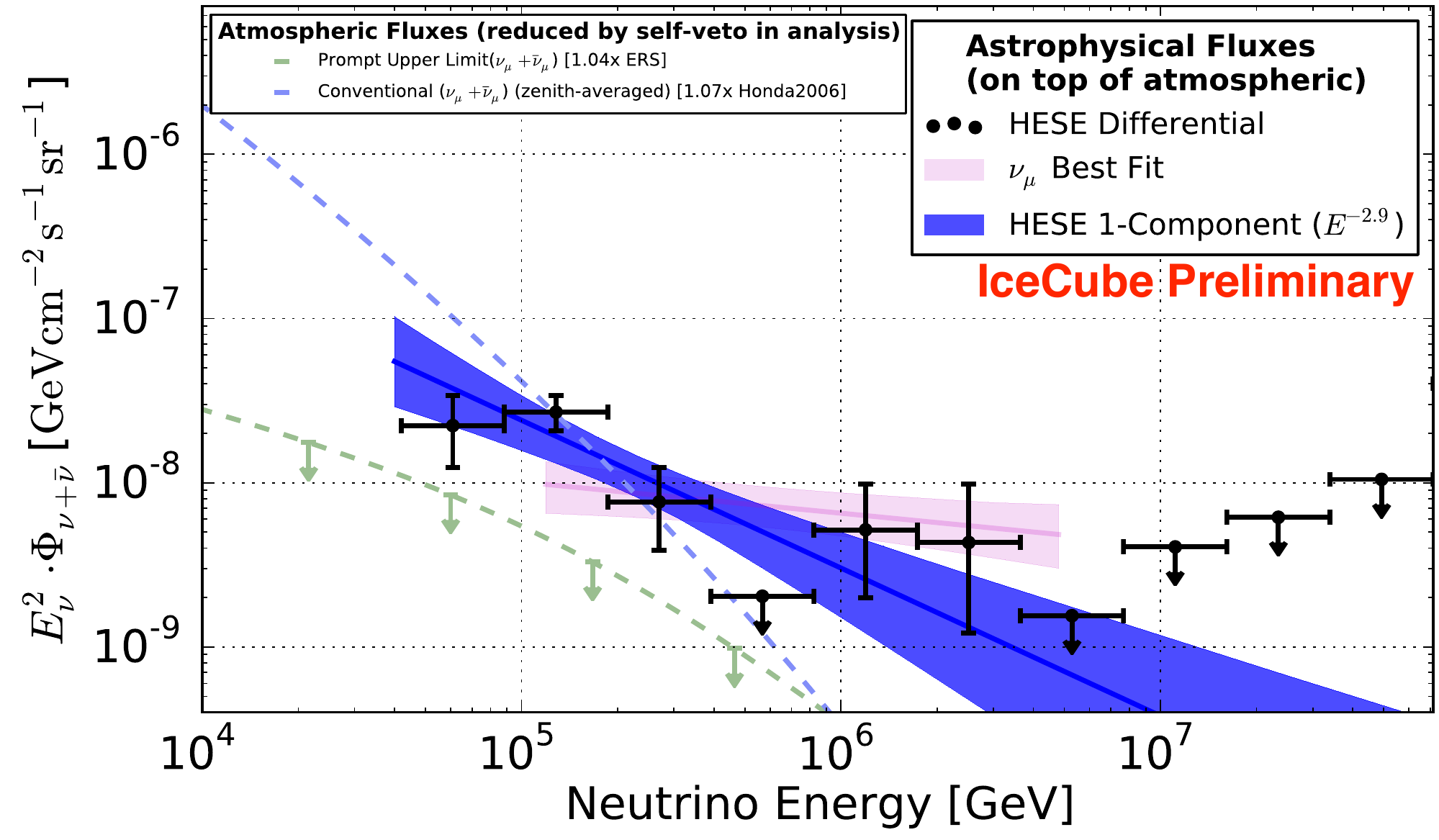}
\vspace{0.3cm} 
\includegraphics[width=0.7\textwidth]{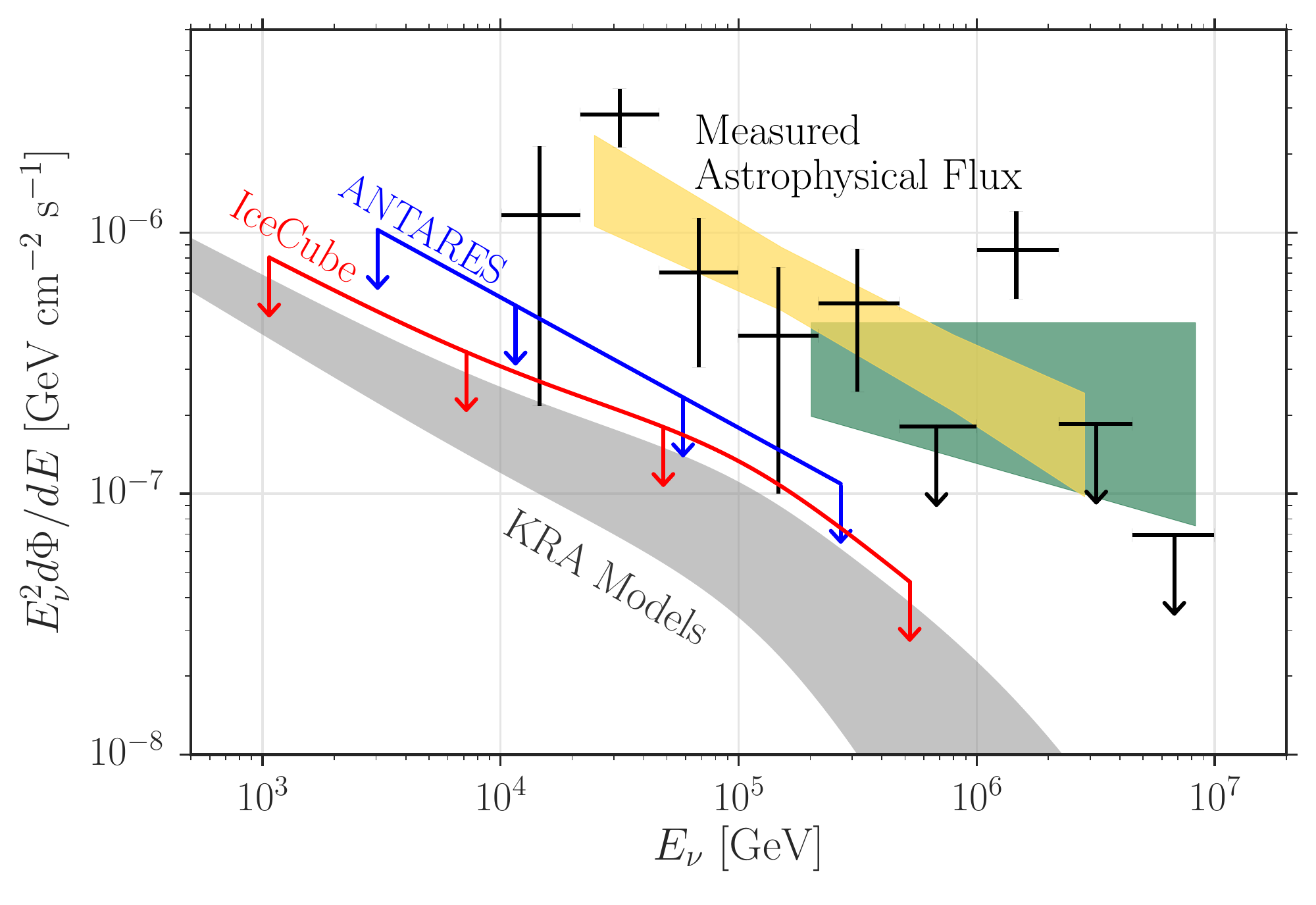}
\vspace{-0.5cm}
  \caption{\small 
{\it Top}: Best-fit of the per-flavor neutrino fluxes as a function of energy. The black points with $1\sigma$ uncertainties are extracted from a combined likelihood fit of all background components together with an astrophysical flux component with an independent normalization in each band (assuming an $E^{-2}$ spectrum within each band and atmospheric neutrino and muon fluxes subtracted). The best-fit conventional flux and the upper limit for prompt neutrinos are shown separately, not taking into account the HESE self-veto which actually reduces their contribution.
The blue band shows the $1\sigma$ uncertainties of a single power-law fit to the HESE data. The pink band shows the fit for the muon neutrino data, again with $1\sigma$ uncertainties. Its length indicates the approximate range providing 90\% of the significance of this analysis \cite{ICRC-Kopper}.
{\it Bottom}:
Upper limits on the three flavor neutrino flux from the Galaxy with respect to KRA model predictions \cite{KRA} and the measured astrophysical flux \cite{ICRC-Galacticplane}. Dots, yellow and green bands have the same meaning as the bands in the top figure.
}
\label{l-spectrum-combined}
\end{figure}

While both analyses (HESE and through-going muons)  have reached a significance for a strong 
non-atmospheric contribution of more than 5\,$\sigma$, the spectral indices 
$\gamma$ of the astrophysical flux from both analyses disagree:  $\gamma = 2.92 \pm 0.33/0.29$ for the HESE events 
(unbroken spectrum $E^{-\gamma}$) and $\gamma = 2.19 \pm 0.10$ \cite{muon-8years} for the throughgoing muons. Adding two more years to the HESE sample has resulted in an even softer energy spectrum since all events of the recent two years have energies below 200 TeV.  Fig.\,\ref{l-spectrum-combined}a 
shows the two fits under the assumptions of a single-power law. The possibility that all but the three PeV HESE events emerge from pion/Kaon/charm decays in the atmosphere is excluded by the zenith angle distribution.
 
In \cite{combined} the flavor ratio of the astrophysical neutrino flux has been investigated. It is consistent with an observed flavor ratio $\nu_e : \nu_{\mu} : \nu_{\tau} = 1 : 1 : 1$ and also with source neutrino ratios 1:2:0 (pion decay) and 0:1:0 (pion decay with suppressed muon decay) while largely excluding 1:0:0 (neutrinos from neutron decay).

The hope to see any clustering of the HESE and muon-track events at highest energies has not fulfilled. An initial indication of clustering of HESE events close to the Galactic center has vanished with more statistics.
In addition, ANTARES has looked to a point source at the position of IceCube's initial excess and could
exclude that it is due to a point source, assuming that the extension of the source does not exceed 
0.5 degrees
and that the spectrum follows an $E^{-2}$ shape \cite{Antares-check}.

A recent IceCube analysis has used 7 years of the medium-energy $\nu_{\mu}$ data (which are optimized to search for point sources, see next section) to set constraints on the diffuse emission of neutrinos from the Galactic plane  \cite{ICRC-Galacticplane}. The resulting limits are shown in 
Fig.\,5b
and compared to the  flux of the HESE and highest-energy $\nu_{\mu}$ data. They exclude that more than 14\% of the observed diffuse astrophysical flux come from the Galactic plane. However, the limit is not far from model predictions (gray band). Joining IceCube and ANTARES data and exploiting cascade-like events in addition to the $\nu_{\mu}$ sample may drive the sensitivity into the region predicted by KRA models.


\vspace{-0.4cm}



\subsection{Search for steady point sources}

For the standard steady-source search, a sample of through-going muons with good angular resolution
(median error smaller $1\deg$) is selected. In the lower hemisphere, the Earth acts as filter against
muons generated in the atmosphere. In the upper hemisphere, a radical energy cut removes most of the
atmospheric muons which have a rather soft energy spectrum, but naturally also rejects all but the most energetic cosmic neutrinos. Therefore only hard-source spectra would result in a significant number of events from the upper hemisphere (for IceCube: South).

\begin{figure}[h]
\centering
\includegraphics[width=0.85\textwidth]{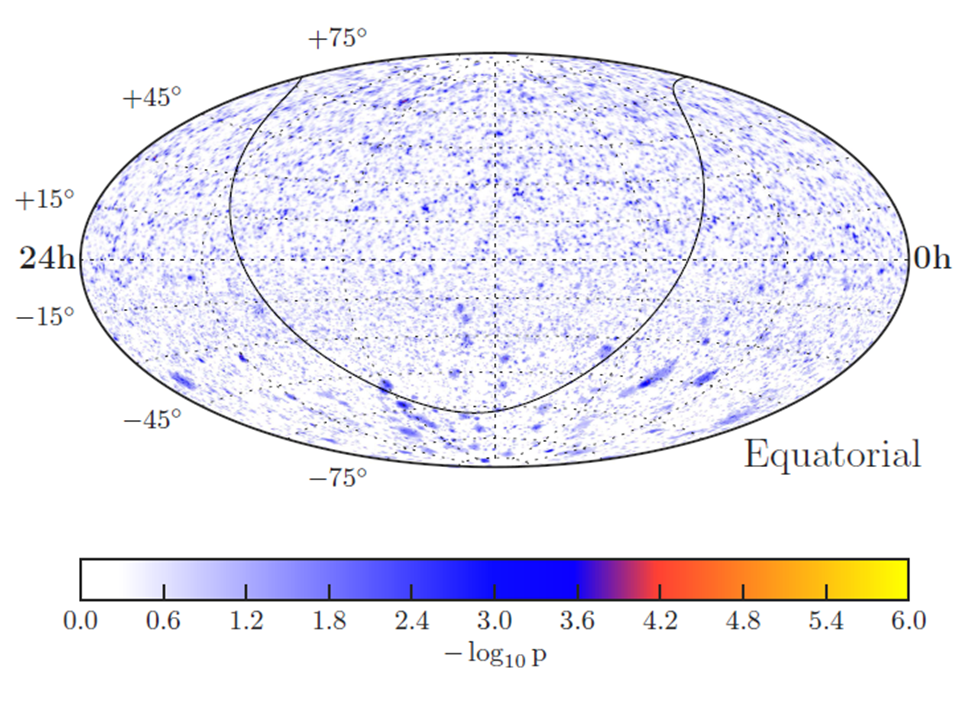}
\vspace{-0.1cm}
  \caption{\small
	All-sky plot of seven years IceCube data in equatorial coordinates. Shown is the
	negative logarithm of the pre-trial p-value, assuming no clustering as null-hypothesis 
	\cite{point-sources}.	
	}
\label{l-skyplot7years}
\end{figure}

Figure \ref{l-skyplot7years} shows the all-sky plot of seven years 
data, with 422 791  upward muons from neutrino interactions
and 289 078 downward muons, the latter almost all from atmospheric showers. The downward sample 
contains also 961 tracks starting inside the detector, i.e. generated in neutrino interactions \cite{point-sources}.

\begin{figure}[h]
\centering
\includegraphics[width=0.8\textwidth]{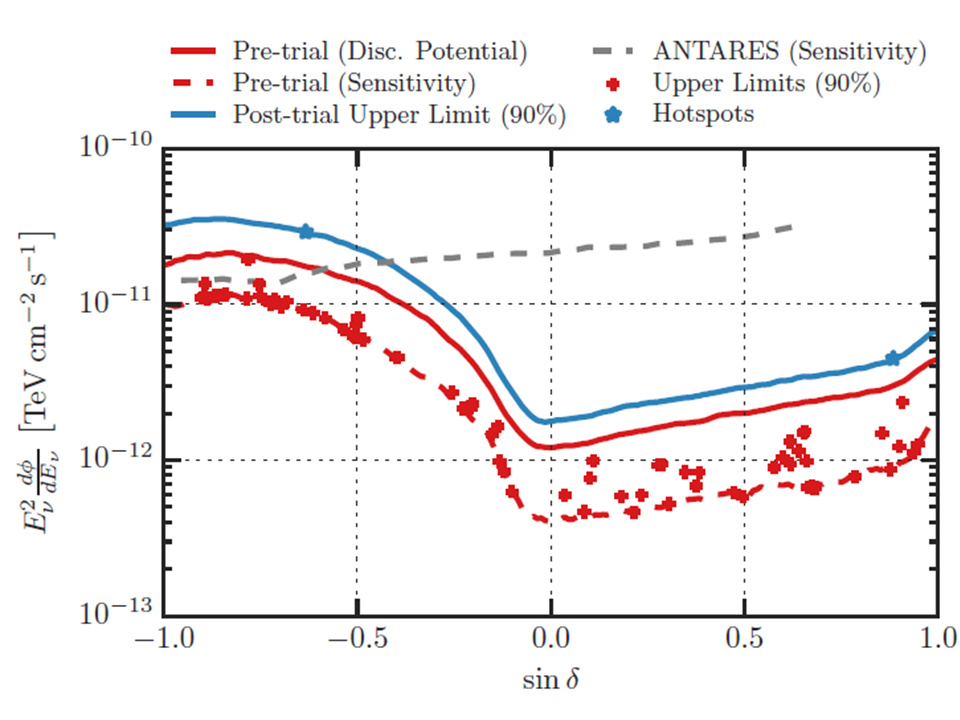}
	\vspace{-0.1cm}
	\caption{\small
	Discovery potential and sensitivity (red solid and dashed, respectively) versus declination, assuming an unbroken
	$E^{-2}$ neutrino spectrum. Upper limits of 32 pre-selected source candidates
	are given as red crosses, the blue line represents the upper limit for the most significant spots in each half of the sky (actual positions of the spots are given by blue stars). The gray line shows the results from ANTARES.
	See \cite{point-sources} for details.
	}
\label{l-pt-limits}
\end{figure}

No significant excess is found, resulting in the flux constraints show in Figure \ref{l-pt-limits}. Apart from sensitivities and limits for selected sources, the discovery potential is shown, i.e. the flux that would lead to a $5\sigma$ discovery of a source in 50\% of the cases. 

\begin{figure}[h]
\vspace{-0.1cm}
\centering
\includegraphics[width=0.75\textwidth]{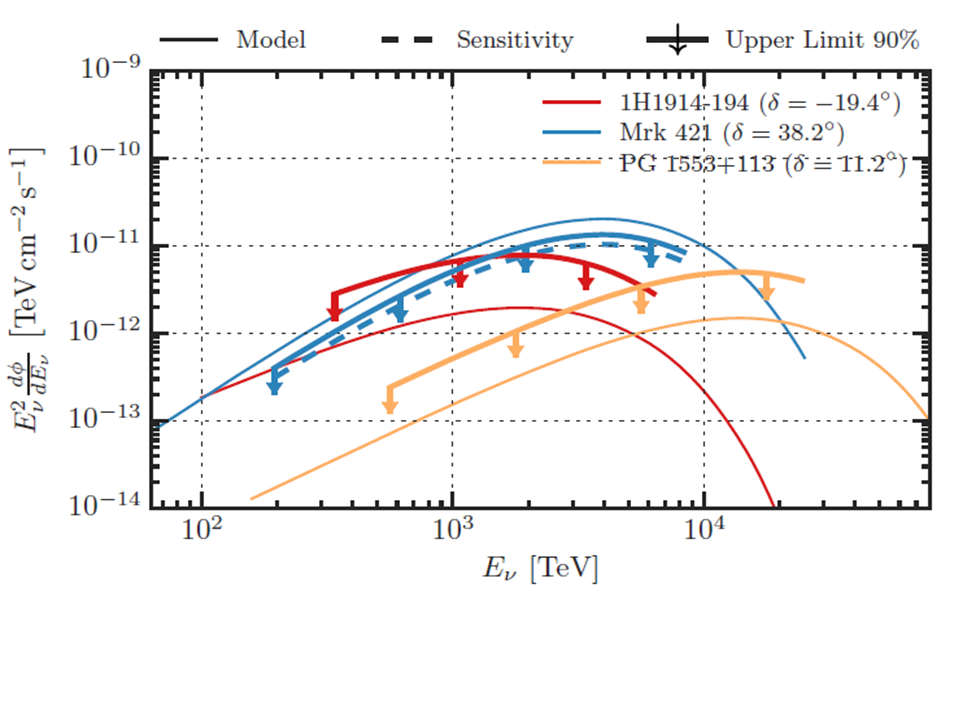}
\vspace{-1cm}
\caption{\small
Differential energy spectra versus neutrino energy for blazars of the BL Lac type compared to model predictions
\cite{Petropoulou}. Thick lines give the 90\% upper limits from IceCube, thin lines represent the model. The sensitivities of the Icecube search are shown as dashed line. 90\% upper limit and sensitivity are shown for the energy interval where 90\% the events originate that are most signal-like \cite{point-sources}.
}
\label{l-limitsblazars}
\end{figure}

One can then compare these values to predictions for selected sources. 
Fig.\,\ref{l-limitsblazars} compares our sensitivities and the obtained 90\% upper limits to predictions \cite{Petropoulou} for three blazars. The limits are within a factor 5 of the predictions, for Mkr 421 even slightly below predictions. Similar relations hold for the Crab nebula -- always optimistically assuming that the gamma flux observed from these sources is basically due to $\pi^0$ decay and not to inverse Compton scattering. 

From these figures one could conclude that an improved angular reconstruction and twice more data could bring us close to discovery. For blazars, however, this hope is downsized by various 
blazar stacking analyses, none of them yielding an excess in the directions of blazars. The most recent one \cite{blazarstacking} indicates that only 4-6\% of the observed diffuse astrophysical muon neutrino flux could come from blazars.

\subsection{Search for transient sources}

To improve the signal-to-background ratio one can search for transient signals, preferentially
in coincidence with an observation in electromagnetic waves. Examples are flares of Actice Galactic Nuclei (AGN) or Gamma-Ray Bursts (GRB). GRBs are interesting objects since there 
are models which assume that they are the dominant source of
the measured cosmic-ray flux at highest energies, either by neutron escape \cite{Ahlers}
or by escape of both neutrons and protons \cite{WB-GRB} from the relativistic fireball. Naturally models where protons a kept in the acceleration region and only neutrons escape and constitute the observed cosmic ray flux give a higher neutrino/cosmic ray ratio.

\begin{figure}[h]
\centering
\includegraphics[width=0.65\textwidth]{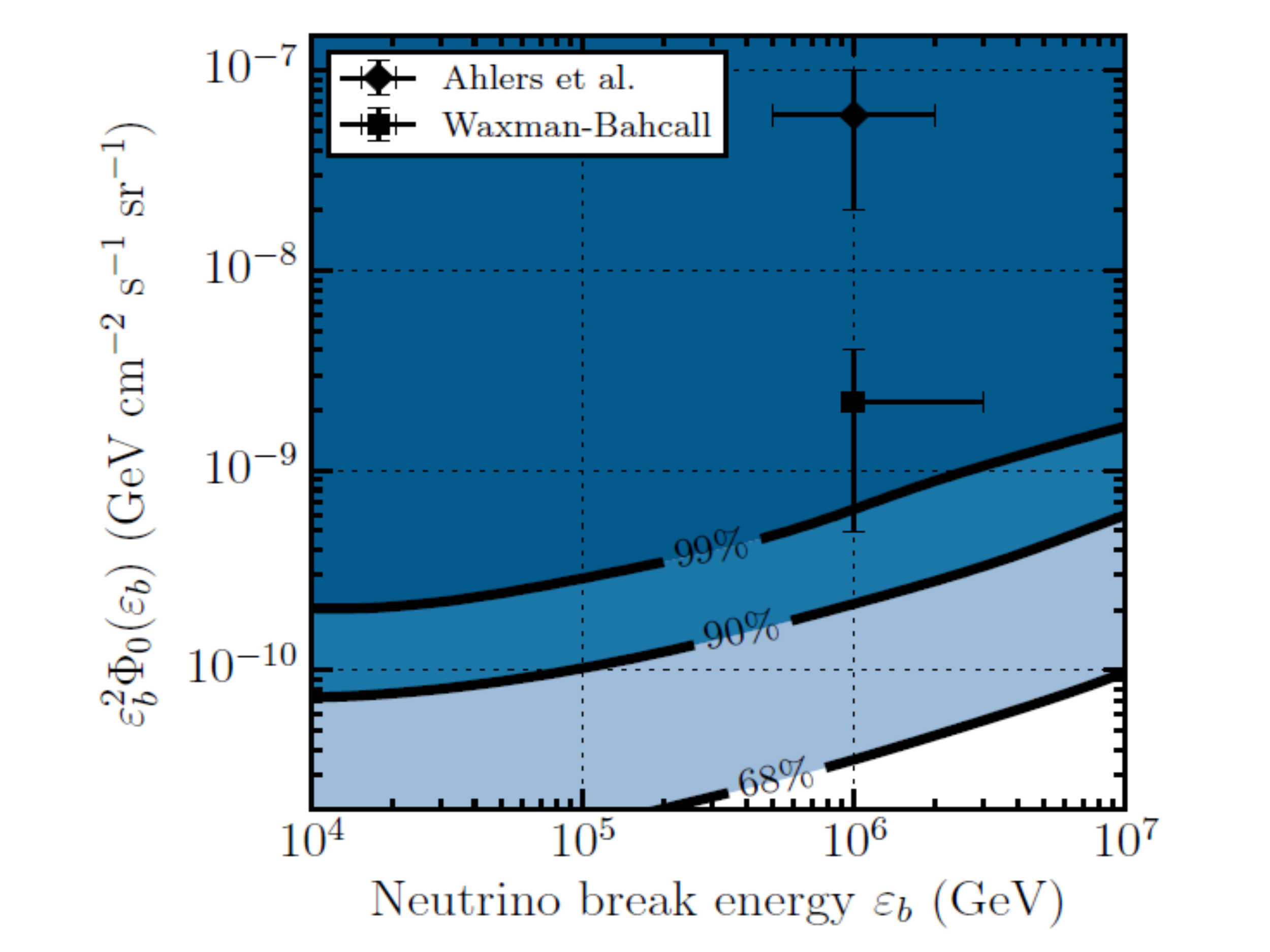}
  \caption{\small
Excluded regions for 99\%, 90\% and 68\% confidence level of the generic
double broken power law neutrino spectrum as a function of the
first break energy $\epsilon_b$ and per-flavor quasi-diffuse flux normalization
derived from upward and downward muon tracks and all-sky cascades.
	}
\label{l-GRB}
\end{figure}

All three collaborations -- Baikal, ANTARES and IceCube -- have searched for neutrinos in local and spatial
coincidence with GRBs.  In particular
IceCube limits on neutrinos from GRBs have drastically improved over the recent years. A recent analysis has combined the searches for spatial and temporal coincidences
of upward and downward tracks and cascade-type events with 1172 GRBs. No  significant correlations between the gamma-ray signals and neutrinos have been observed. Figure \ref{l-GRB} shows exclusion contours
for double broken power-law spectra, with breaks from $E^{-1} \times \epsilon_b$ to $E^{-2}$ at 
energy $\epsilon_b$, and from $E^{-2}$ to $E^{-4} \times (10\epsilon_b)^2$ at an 
energy $10\epsilon_b$.

Both models, those with cosmic ray escape via neutrons and those which
allow additionally for cosmic ray escape via protons,
are excluded at over 90\% confidence level, with most of
the model assumption phase space excluded at over the
99\% confidence level, greatly constraining the hypothesis that GRBs
are significant producers of ultra-high energy cosmic rays in the prompt GRB
phase.

30 years after the discovery of supernova SN1987 it is worth to highlight that IceCube
runs a supernova trigger, with a duty time of more than 99\%. The trigger reacts to
a collective rise in photomultiplier counting rates on top of the dark-noise rate.
This rise would be due to the feeble signals
from $\nu_e$
reactions close to a photomultiplier. A 1987A-type supernova
at 30 kpc distance (edge of the Galaxy) would lead to an collective-rate enhancement with
a significance of about 20 standard deviations, and even at distance of the Large Magellanic Cloud
(50 kpc) the excess would reach $6-7\sigma$ \cite{SN}. IceCube is part of the SuperNova Early Warning
System (SNEWS \cite{SNEWS}), together with the underground neutrino detectors Borexino, Super-Kamiokande,
LVD and Kamland which, in the case of a significant coincidence from more than one of the detectors,
would alarm the astronomers community. However, no significant neutrino signal has been recorded yet, 
neither with the analogous trigger of IceCube's predecessor AMANDA nor with that of IceCube.

\subsection{Real-time alert and follow-up programs}

With no steady sources of high-energy neutrinos observed so far, neutrinos produced
during transient astrophysical events are a viable alternative. High-energy neutrinos from the
prompt phase of GRBs or MeV neutrinos from a supernova collapse as discussed in the previous
section are just two examples. Coincident detections could enhance the significance of the
IceCube observation and, more generally, contribute to the mosaic of informations from different
messengers, providing a more complete picture of the source. Since IceCube and ANTARES have nearly
$4\pi$ acceptance (depending on energy), they could could trigger detections with pointing
devices like optical or gamma-ray telescopes, which otherwise would have been missed.
 
Both collaborations run a number of high-energy alert and follow-up programs \cite{alerts,alertsAnt} 
which react to particular single events. 
In the case of IceCube, neutrino alert candidates are identified in real-time at the South Pole. A brief message sent to
the North is automatically issued to the Gamma-Ray Coordinates Network (GCN \cite{AMON}) via the
Astrophysical Multimessenger Observatory Network (AMON \cite{AMON}). In parallel, quality checks are
applied and the directional and energy reconstruction refined. Results from that are completed
within a few hours and lead to an updated alert notification in the form of a CGN circular.
IceCube runs two of these alerts: a ``HESE Track alert'' which is issued if a track-like HESE event is recorded
(4.8 events expected per year, with 1.1 being of astrophysical origin) and 
an ``EHE Track Alert''
which is based on a selection which originally targets cosmological neutrinos (10 PeV to 1 EeV) but 
here is modified to be sensitive down to 500 TeV (about 5 alerts per
year). 

Apart from these public alerts, IceCube also issues alerts 
to optical, X-ray and gamma-ray observatories which are based on neutrino {\it multiplets}. 
These alerts are based on individual agreements with these observatories. 
The multiplets can be due to phenomena on the second-to-minute scale (high-energy neutrinos from relativistic jets in SN or GRB), or to phenomena of the hour-to-week scale
(like AGN flares).  
None of the alerts yet has led to a significant correlation, although at least two cases have generated some 
initial excitement.
The one \cite{doublet} was a neutrino doublet detected in March 2012 which triggered follow-up observations by the Palomar Transient Factory (PTF). PTF found a Type IIn supernova within an error radius of $0.54\deg$ of the direction of the doublet. A Pan-STARRS1 survey, however, showed that its explosion time was
at least 158 days before the neutrino alert, 
so that a causal connection is unlikely.
The second case \cite{triplet} was the first triplet: three neutrinos arriving within 100 s of
one another at February 17, 2016. Follow-up observations by SWIFT's X-ray telescope, by ASAS-SN, LCO and MASTER at optical wavelengths, and by VERITAS in the very-high-energy gamma-ray regime did not detect any likely electromagnetic counterpart. In a refined reconstruction, the directions of the events changed 
slightly, so that the triplet turned to a double-doublet (error circle of the one event overlapping with those of the two others, but not all three with each other).
Still, these two cases impressively illustrate the potential of and challenges for future follow-up campaigns. Although no significant correlations have been detected so far, the IceCube/ANTARES alerts and the triggered electromagnetic-domain observations herald the era of multi-messenger observation. This remark also applies to the follow-up programs where IceCube scrutinizes its own data to search for correlations with signals from Gravitational Waves \cite{GW}.

\section{Where do we go?}

Four years after the detection of cosmic neutrinos, we have learned a lot about their
spectrum and flavor composition. 
We have learned that blazar jets and GRBs can contribute only a small fraction to
the observed astrophysical neutrino flux. The spectral features of this flux 
(single power law or two power law) open new questions about the contributing source classes.
No individual sources have been detected yet. The non-observation of neutrinos coinciding with GRBs strongly constrains models which attribute the highest-energy cosmic rays to GRBs. Neutrino events possibly related to supernova explosions have been observed, although with a non-negligible probability for a chance occurrence. 
No neutrinos have been observed that could be attributed to the GZK effect \cite{BZ}, but the non-observation starts constraining evolution scenarios
for ultra-high energy cosmic rays sources (not addressed in this report).

IceCube continues collecting data. A twofold statistics combined with improved directional precision,
also for cascade-like events,
and better understanding of systematics effects will considerably improve the understanding of what has
been observed so far and may even provide first detection of individual (point-like or extended) sources.
IceCube's capabilities, however, are limited by its size, the chance to detect neutrinos from the central part of our Galaxy are constrained by its location at the South Pole.

The next important steps are being done at the Northern hemisphere:  GVD in Lake Baikal and KM3NeT in the Mediterranean Sea.

\newpage

\subsection{Baikal-GVD}

Baikal-GVD (Gigaton Volume Detector) is configured in ``clusters'', where each cluster consists of eight strings, instrumented over a length of 520\,m with 36 optical modules. The OMs house 10'' Hamamatsu photomultipliers with a high-sensitive photocathode. After an extensive period of in-situ tests of single components and prototype strings, a first cluster with 24 OMs per string was deployed in Spring 2015. One year later, it was upgraded to a full cluster with 36 PMs per string, and in Spring 2017 a second cluster was added.
First preliminary results of the 2016 cluster have been presented at the ICRC 2017 \cite{Baikal-ICRC}.

\begin{figure}[h]
\centering
\includegraphics[width=0.8\textwidth]{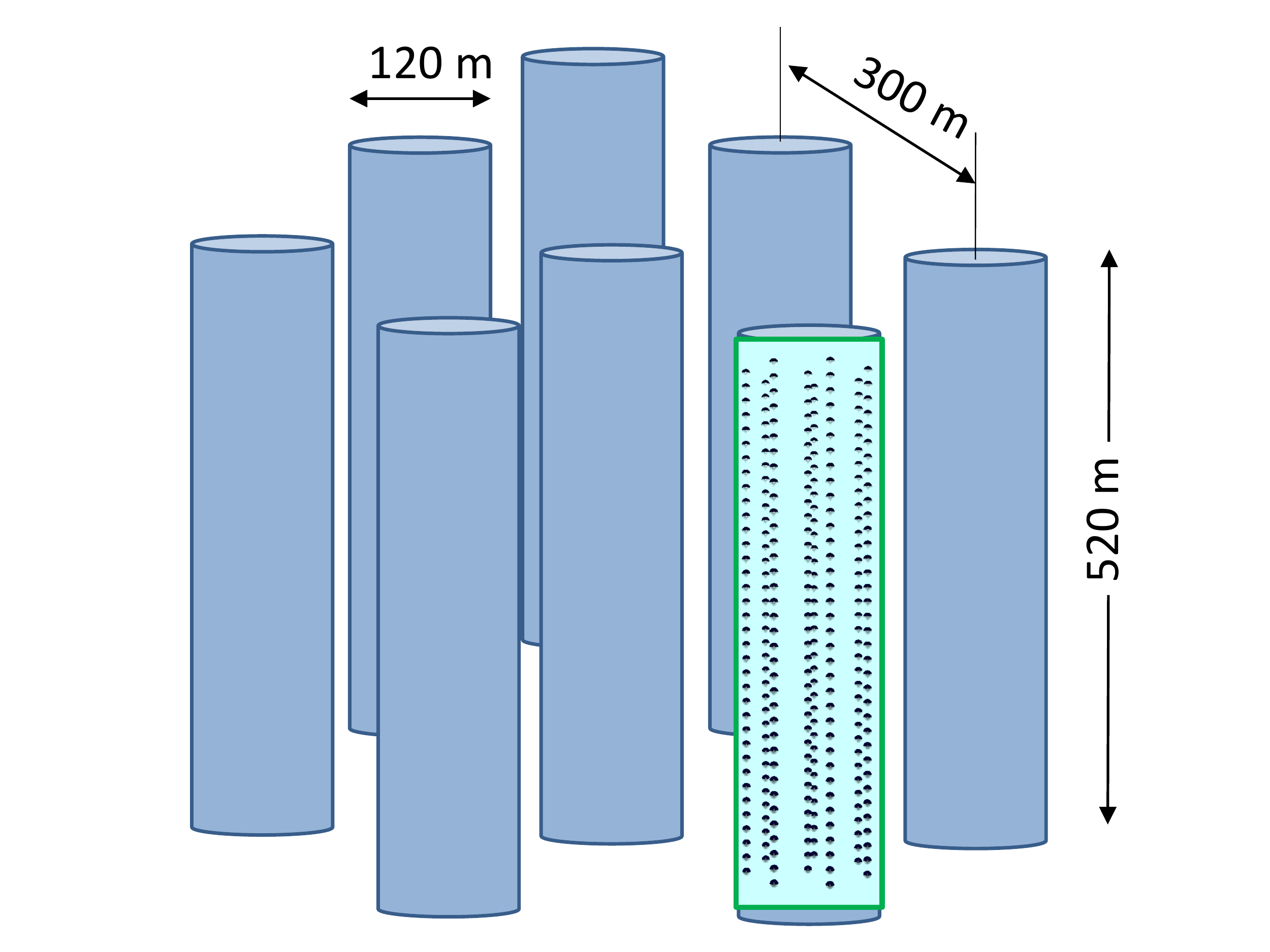}
  \caption{\small
Schematic view of phase-1 of Baikal-GVD, consisting of 8 clusters, each with 120\,m diameter and 520\,m height.
A cluster consists of eight strings with 36 optical modules along each string.  
	}
\label{GVD}
\end{figure}

Baikal-GVD will be built in two phases. Phase-1 will consist of eight clusters, each 120\,m in diameter,
with lateral distances of 300\,m (see Fig.\,\ref{GVD}). The effective volume for cascades in the 10-100 TeV energy range will be about 0.4\,km$^3$, the sensitivity to muons is negligible below 1\,TeV but rapidly raises in the multi-TeV range. Phase-1 is financed and planned to be completed in 2020/21. In a second phase, Baikal-GVD will be extended to 18 clusters and then surpass the cubic kilometer benchmark.  

\subsection{KM3NeT}

\begin{figure}[h]
\centering
\includegraphics[width=0.65\textwidth]{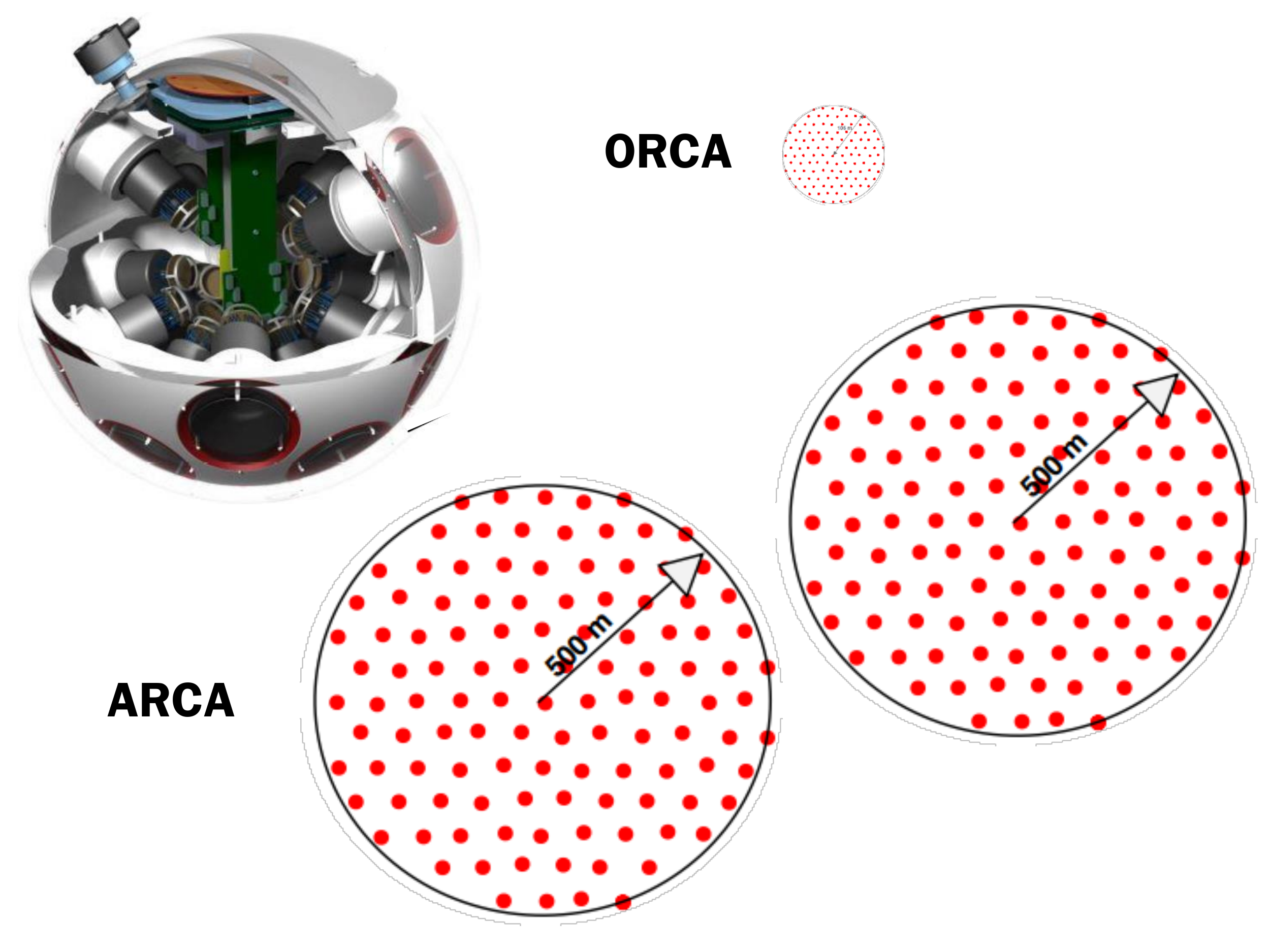}
  \caption{\small
The two incarnations of KM3NeT. The two ARCA blocks (bottom) have diameters of 1\,km and a height of about 600\,m and focus to high-energy neutrino astronomy. ORCA (top) is a shrinked version of ARCA with only 200\,m diameter and 100\,m height. Both ARCA and ORCA have 115 strings with 18 optical modules (OMs) per string. Top left, a drawing of an
OMs is shown. Each OM houses 31 small photomultipliers.
	}
\label{KM3NeT}
\end{figure}

KM3NeT has two main, independent objectives: a) the discovery and subsequent observation of
high-energy cosmic neutrino sources and b) precise oscillation measurements and the determination of the mass hierarchy of neutrinos. For these purposes
the KM3NeT Collaboration plans to build an infrastructure distributed over three sites:
off-shore Toulon (France), Capo Passero (Sicily, Italy) and Pylos (Peloponnese, Greece).
In a configuration to be realized until 2020/22, KM3NeT will consist of three so-called building blocks
(``KM3NeT Phase-2''). 
A building block comprises 115 strings, each string with 18 optical modules.Two
building blocks will be sparsely configured to fully explore the IceCube signal with a comparable instrumented
volume, different methodology, improved resolution and complementary field of view, including the Galactic
plane. These two blocks will be deployed at the Capo Passero site and are referred to as ARCA: Astroparticle Research with Cosmics in the Abyss. The third building block will be densely configured to precisely measure atmospheric neutrino oscillations.
This block, being deployed at the Toulon site, is referred to as ORCA: Oscillation Research with Cosmics in the Abyss (see Fig.\,\ref{KM3NeT}). 

A novel concept has been chosen for the KM3NeT optical module:
The 43\,cm glass spheres of the DOMs will be equipped with 31 PMTs of 7.5\,cm
diameter, with the
following advantages: a) The overall photocathode area exceeds that of a 25\,cm
PMT by more than a factor three; b) The individual readout of the PMTs results
in a very good separation between one- and two-photoelectron signals which is
essential for online data filtering; c) some directional information is
provided. This technical design has been validated with in situ prototypes. A
cross-sectional view of this DOM is shown at the top of Fig.\,\ref{KM3NeT}.

\vspace{0.3cm}
With a fully equipped ARCA, IceCube's cosmic neutrino flux could be detected
with high-significance within one year of operation. In practise the detector will
be deployed in stages allowing to reach the one-year sensitivity of two clusters much before
the second cluster is fully installed. Actually the same is true for Baikal-GVD, with a good chance that
GVD will find cosmic neutrinos before ARCA. ORCA could determine the neutrino mass hierarchy with at least 3\,$\sigma$ significance after three years of operation, i.e. as early as 2023. 

\subsection{IceCube-Gen2}

The progress from IceCube over the next decade is limited by the modest
numbers of cosmic neutrinos measured, even in a cubic
kilometer array. In \cite{Gen2} a vision
for the next-generation IceCube neutrino observatory is presented. 
At its heart is an expanded array of optical modules
with a volume of 7 to 10\,km$^3$.  This high-energy array will mainly address
the 100\,TeV to 100\,PeV scale. It has
the potential to deliver first
GZK neutrinos, of anti-electron neutrinos produced via
the Glashow resonance, and of PeV tau neutrinos,
where both particle showers associated with the production
and decay of the tau are observed (``double bang events''). 

\begin{figure}[h]
\centering
\includegraphics[width=0.8\textwidth]{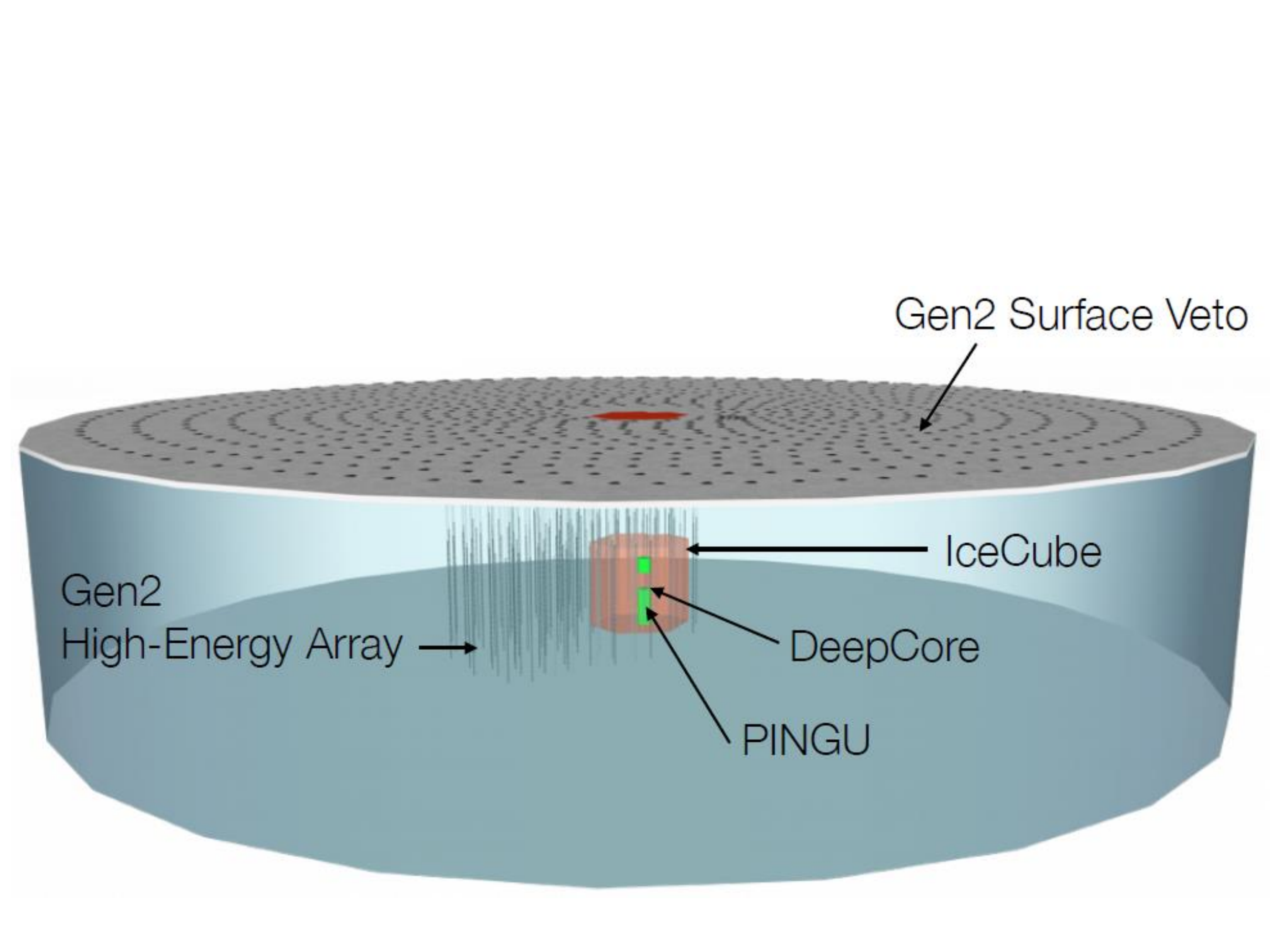}
  \caption{\small
Schematic view of IceCube Gen-2, comprising the existing IceCube array with
its densely equipped inner region DeepCore, the high-energy array of Gen2, the super-densely
equipped PINGU sub-detector, and an extended surface array. Not shown is the radio array ARA with 
its size exceeding that of the basic surface array.
	}
\label{Gen2}
\end{figure}

Another possible component of IceCube-Gen2 is the 
PINGU sub-array. It targets -- similar to ORCA -- precision measurements
of the atmospheric oscillation parameters and the
determination of the neutrino mass hierarchy. The facility's
reach would further be enhanced by exploiting the
air-shower measurement and vetoing capabilities of an
extended surface array. Moreover, a radio array 
(``ARA'', for Askarian Radio Array) will achieve improved
sensitivity to neutrinos in the $10^{16} - 10^{20}$ eV energy
range, including GZK neutrinos. Figure \ref{Gen2} sketches a possible design of IceCube-Gen2.

\newpage

For point sources, the high-energy array will have five times better
sensitivity than IceCube, and the rate for events at energies above a few hundred TeV will be ten times higher than for IceCube.

\subsection{Expectations}

Personal expectations do not necessarily agree with the optimistic time schedules of the experiments. 
Still, one can certainly expect that at the early 2020s, the IceCube cosmic neutrino signal will be scrutinized by KM3NeT-ARCA and Baikal-GVD, with different experimental systematics and from the Northern hemisphere: what is the exact form of the spectrum, what is the flavor composition, what is the contribution from our Galaxy? The diffuse flux from the Galactic plane will be almost certainly discovered, at least if moderately conservative predictions are correct. The discovery of point or extended sources in our Galaxy is not guaranteed but seems likely. Some bet that the first discovery of individual sources will come from transient events correlated to electromagnetic or gravitational wave
observations. Actually, within the next few years, this seems to be the most promising chance for IceCube. 

Another aspect of increased discovery potential are analyses combining data from different detectors. Within the Global Neutrino Network GNN \cite{GNN}, such analyses are being performed between IceCube and ANTARES \cite{common} and bear the chance to detect diffuse emission from the Galaxy rather soon. When Baikal-GVD and ARCA approach the cubic kilometer scale, these efforts will become even more important than now.

With the appearance of four or more ARCA blocks, with GVD Phase-2 and with IceCube Gen2 in the second
half or at the end of the 2020s,
one could have 3--5 km$^3$ instrumented volume in the North and 7--10 km$^3$ in the South.
I seems unlikely that this coordinated attack to the high-energy neutrino frontier would
fail to detect structures and individual sources. That, at the end, will allow charting the 
neutrino landscape to which IceCube has enabled a first glance!


\begin{thebibliography}{99}

{\small

\bibitem{Spiering-History} C.\,Spiering, Eur.\,Phys.\,J. H37 (2012) 515.

\bibitem{Science-2013} M.G.\,Aartsen et al. (IceCube Coll.) Science 342 (2013) 1242856.

\bibitem{KM3NeT} S. Adri\'an-Mart\'\i nez et al.\ (KM3NeT Coll.), J.\,Phys. G43 (2016) 084001.

\bibitem{Baikal} A.\,Avronin et al.\ (Baikal Coll.), Phys.\,Part.\,Nucl.~46 (2015) 211.

\bibitem{Gen2}  M.G.\,Aartsen et al. (IceCube Coll.), arXiv:1412.5106.

\bibitem{point-sources} M.G.\,Aartsen et al. (IceCube Coll.) Astrophys.\,J.\, 835 (2017) no.2, 151.


\bibitem{Baikal-atm-Balkanov-1999} R.\,Balkanov, et al. (Baikal Coll.),
Astropart.\,Phys.\,12 (1997) 75.

\bibitem{Igor-Christian-skyplot} I.\,Belolaptikov and C.\,Spiering, Baikal-Amanda Internal report, 2005.

\bibitem{Antares-results} P.\,Coyle and C.W.\,James, arXiv:1701.02144.

\bibitem{IceCube}  M.G.\,Aartsen et al. (IceCube Coll.), JINST 12,\,3 (2017) P03012. 
  
\bibitem{Lipari} P.\,Lipari, Nucl.\,Instrum.\, Meth.\, A, 567 (2006) 405.

\bibitem{Ernie} M.G.\,Aartsen et al., Phys.\,Rev.\,Lett.\, 111 (2013) 021103. 

\bibitem{ICRC-Kopper} C.\,Kopper for the IceCube Coll., ICRC 2017.

\bibitem{muon-7years} M.G.\,Aartsen et al. (IceCube Coll.) Astrophys.\,J.\,833(2016)\,3.

\bibitem{muon-8years} C.\,Haack and C.\,Wiebusch for the IceCube Coll., ICRC 2017.

\bibitem{combined} M.G.\,Aartsen et al. (IceCube Coll.) Astrophys.\,J.\,808(2015)\,98.

\bibitem{Antares-check} S. Adri\'an-Mart\'\i nez et al. (ANTARES Coll.) Astrophys.\,J.\,786 (2014) L5.

\bibitem{KRA} D.\,Gaggero, D.\,Grasso, A.\,Marinelli, A.\,Urbano and M.\,Valliet, 
Astrophys.\,J.\,815,2 (2015) L25.

\bibitem{ICRC-Galacticplane} C.\,Haack and J.\,Dumm for the IceCube collaboration, ICRC 2017.

\bibitem{Petropoulou} M.\,Petropoulou, S.\,Dimitrakoudis, P.\,Padovani,
A.\,Mastichiadis and \\ E.\,Resconi, Mon.\,Not.\,Roy.\,Astron.\,Soc., 448 (2015) 2412.

\bibitem{blazarstacking} M.\,Huber and K.\,Krings for the IceCube Collaboration, ICRC 2017.

\bibitem{GRB-2017} M.G.\,Aartsen et al. (IceCube Coll.), Astrophys.\,J.\,843 (2017) no.2, 112.

\bibitem{Ahlers} M.\,Ahlers, M.\,Gonzalez-Garcia and F.\,Halzen, Astropart.\,Phys.\,35 (2011) 87.

\bibitem{WB-GRB} E.\,Waxmann and J.\,Bahcall, Phys.\,Rev.\,Lett.\,78 (1997) 2292.

\bibitem{SN} R.Abbasi et al. (IceCube Coll.) Astron.Astrophys. 535 (2011) A109.

\bibitem{SNEWS} {\it http://snews.bnl.gov/}

\bibitem{alerts} E.\,Blaufuss for the IceCube collaboration, ICRC 2017, and M.G.Aartsen et al. (IceCube Coll.) Astrophys.\,J.\,811 (2015) no.1, 52. 

\bibitem{alertsAnt} S. Adri\'an-Mart\'\i nez et al. (ANTARES, TAROT, ROTSE, Swift and Zadko 
Collaborations), JCAP 1602, 2 (2016) 062. 

\bibitem{AMON} see {\it https://gcngsfc.nasa.gov/} \hspace{3mm} and \hspace{3mm} {\it http://amon.gravity.psu.edu/}

\bibitem{doublet} M.G.\,Aartsen et al. (IceCube, PTF and SWIFT Coll.), Astrophys.J. 811 (2015) no.1, 52.

\bibitem{triplet} M.G.\,Aartsen et al. (IceCube Coll.) arXiv:1702.06131.

\bibitem{GW} M.\,Albert et al. (ANTARES and IceCube Coll.) arXiv:1703.06298.

\bibitem{BZ} V.\,Berezinsky and G.\,Zatsepin, G., Yad.\,Fiz. 11 (1970) 200.

\bibitem{Baikal-ICRC} Zh. Djilkibaev for the Baikal Collaboration, ICRC 2017. 

\bibitem{GNN} {\it http://www.globalneutrinonetwork.org/ }

\bibitem{common}  S. Adri\'an-Mart\'\i nez et al.,
Astrophys.\,J. 823, 1 (2016) 65. 

}

\end{thebibliography}
\end{document}